\documentclass{ws-ijgmmp}

\begin{document}

\markboth{Partha Sarathi Debnath}
{Bulk Viscous Cosmological model  in $f(R, T)$ theory of gravity}

\catchline{}{}{}{}{}
%
%%%%%%%%%%%%%%%%%%%%%%%%%%%%%%%%%%%%%%%%%%%%%%%%%%%%%%%%%%%%%%%%%%%%

\title{Bulk Viscous Cosmological model  in $f(R, T)$ theory of gravity}

\author{Partha Sarathi Debnath}

\address{Department of Physics, A. B. N. Seal College \\
Coochbehar, West Bengal, Pin-736101, India.\\
\email{parthasarathi6@hotmail.com} }

\maketitle

\begin{history}
\received{(Day Month Year)}
\revised{(Day Month Year)}
\end{history}

\begin{abstract}
In this paper, we have presented bulk viscous cosmological model of the universe in the  modified gravity theory in which the Lagragian of the gravitational action contains a general function $f(R, T) $, where $R$ and $T$ denote the curvature scalar and the trace of the energy-momentum tensor respectively, in the framework of a flat Friedmann-Robertson-Walker model  with isotropic fluid.   We obtain cosmological solution  in $f(R,T)$ theory of gravity, specially of particular choice   $f(R,T)=R+2\lambda T$, where $\lambda$ is a constant, in the presence of bulk viscosity that are permitted in Eckart theory, Truncated Israel Stewart  theory and Full Israel Stewart theory.  The physical and geometrical properties of the models in Eckart, Truncated Israel Stewart  theory and  Full Israel Stewart theory are studied in detail.  The analysis of the variation of bulk viscous pressure, energy density, scale factor, Hubble parameter and deceleration parameter with cosmic evolution are done in the respective theories.  The models are analyzed by comparison with recent observational data. The cosmological  models are compatible with  observations.
\end{abstract}

\section{Introduction}
Recent data \cite{Riess} from cosmological observations suggest that the universe might be passing through an accelerating phase of evolution.  Although the general theory of relativity is a very successful theory to describe most gravitational phenomena for the evolution of the universe, however,   standard cosmological models with perfect fluids of standard fluid forms, that is radiation, matter, stiff matter fail to address properly the late-time acceleration of the universe. The late-time acceleration \cite{Padma} problem along with the dark matter problem are the most difficult challenges to modern gravitational theory.  The late-time acceleration  is attributed to a negative pressure fluid dubbed as dark energy \cite{Cope}, but what is the dark energy for the moment we have absolutely no idea.  The estimate from the recent cosmological observations, PLANCK Collaboration 2015  classifies dark energy which plays a significant role to drive the acceleration, as  consisting of  $\sim $ 69.4$\%$ of the total energy, dark matter contributes $\sim $ 25.8 $\%$ and baryonic matter consists of $\sim $ 4.8$\%$ only. Quite a few theoretical proposals  \cite{Nojiri11} came up to understand the exact fundamental nature of the dark energy. Phenomenological  models \cite{starobinsky,mukherjee}  appear by modifying gravitational sector and/or matter sector of Einstein’s field equation to study geometrical and physical features of the different phases of the universe. The modified theories of gravity with proper curvature correction are also considered to account for  dark energy.  Literature \cite{Nojiri,rev10,rev11} also discussed the reasons why modified gravity approach is extremely attractive in the applications for late-time acceleration of the universe and to understand the problem of dark energy.
A number of  modified gravity theories,  namely, $f (R)$ (where $R$ being the Ricci scalar curvature) \cite{Nojiri1}, $f(T) $
(where $T$ being the trace of the stress-energy tensor) \cite{Cai,Bamba1}, Horava-Lifshitz \cite{Nishioka} and Gauss-Bonnet
\cite{Li} theories have been recently proposed. Recently, Harko {\it et al.} \cite{harko} have introduced another extension of general relativity (GR) called $f(R, T)$ modified theory of gravity, where gravitational Lagrangian is given by an arbitrary function of $R$ and $T$. In recent times such a modification of Einstein's theory is found to describe some of the observed features relevant for cosmology and astrophysics \cite{debnath1, beesham1}. In $f(R,T)$ gravity, literature \cite{baffou} takes into account an interacting cosmological fluid described by generalized Chaplygin gas with viscosity . Relativistic cosmological solutions and the different phases of the universe with non-causal viscous fluid  are studied \cite{singh1} in $f(R,T)$ gravity theory. In $f(R, T)$ gravity  finite-time future singularities are discussed \cite{houndjo}.  The energy condition in $f(R,T)$ gravity theory by incorporating conservation of energy-momentum tensor is studied in the literature  \cite{chakraborty} and it analyzed that $T$ sector can not be chosen arbitrarily but it has special form.  Cosmological solution in modified  $f(R,T)$ gravity theory with $\Lambda(T)$ gravity has been studied in literature \cite{umesh} also.
 Cosmological solutions relevant in  quantum era  have been studied \cite{xu} in $f(R,T)$ gravity. In $f(R,T)$ gravity theory the behavior of gravitational wave is considered in the literature \cite{alves}.  The stationary scenario between dark energy and dark matter is also studied \cite{rudra} in $f(R,T)$ theory.

 It has been shown that viscosity is one of the significant aspects to study evolution of the universe. A number of dissipative processes 
\cite{misner,zimdahl1} in the early universe guide to departure  from perfect fluid assumptions \cite{pavon,lima}, which permit the presence of  viscosity \cite{rev12,rev16} and is, therefore, important to explore cosmological solution with viscosity \cite{rev13,rev17} to study the different phases of evolution of the universe.  In the early universe, viscosity may originate due to various processes e.g.,  decoupling of matter from radiation during the recombination era, particle collisions involving gravitons and formation of galaxies \cite{barrow}. A non-negligible bulk viscous stress is also  important at late-time evolution of the universe \cite{pavon1,rev15} as predicted by observations.  Eckart \cite{eckart} first formulated a relativistic theory of viscosity. However the theory of Eckart suffers from shortcomings, namely, causality and stability \cite{hiskock}. Subsequently, Israel and Stewart \cite{israel} developed a fully relativistic formulation of the theory which is termed as transient or extended irreversible thermodynamics ( in short $\it{EIT}$) which provides a satisfactory replacement of the Eckart theory. Using the transport equations obtained from  $\it{EIT}$, cosmological solutions are explored in Einstein's gravity \cite{arbab,pradhan,colistete}.  Thus it is important to explore cosmological solutions with barotropic fluid and  viscosity described by the transport equation obtained from ${\it EIT}$  in $f(R,T)$ theory of gravity. 

In this paper we study cosmological solutions  in the presence of the bulk viscosity described by Eckart theory,  Truncated Israel Stewart theory and  Full Israel Stewart theory in $f(R,T)$ theory of gravity. The plan of this paper is as follows: in sec. 2, we give the relevant field equations in $f(R,T)$ theory of gravity. In sec. 3, cosmological solutions are presented. In sec. 4,  observational data fitting are considered to study constraint over model parameters. Finally, in sec. 5, we summarize the results obtained.

\section{Field Equations in $f(R,T)$ gravity theory}
In the $f(R,T)$ theory gravity formalism action is given by \cite{harko}
\begin{equation}
I=\int d^4x\sqrt{-g} \left( \frac{1}{2}f(R,T)+L_m \right),
\end{equation}
 where we consider $ 8\pi G=1, \;  c=1$. The $f(R,T)$ is an arbitrary function of Ricci scalar, $R$, and  $T$ $(=g^{\mu\nu}T_{\mu\nu})$ is the trace of the energy-momentum tensor $T_{\mu\nu}$. $L_m $ is the matter Lagrangian density of the matter field. The energy-momentum tensor of matter \cite{landu} is expressed as  $T_{\mu\nu}=-\frac{2}{\sqrt{-g}} \frac{(\sqrt{-g}\delta L_m)}{\delta g^{\mu\nu}}$. Assuming that the matter Lagrangian density depends only on the metric tensor $g_{\mu\nu}$, not on its derivative, one is led to write $T_{\mu\nu}=g_{\mu\nu}L_m -2 \frac{\partial L_m}{\partial g^{\mu\nu}}$. The variation of gravitational action, given by Eq. (1), with respect to metric tensor $g_{\mu\nu}$ yields
\begin{equation}
f_R R_{\mu\nu}-\frac{1}{2}f(R,T) g_{\mu\nu}+ \left[g_{\mu\nu}\nabla_\mu \nabla^\mu -\nabla_\mu \nabla_\nu \right]f(R, T)
 =T_{\mu\nu}-f_T \left[T_{\mu\nu} + \Theta_{\mu\nu} \right],
\end{equation}
where $f_R$ and $f_T$ denote the derivative of $f(R, T)$ with respect to $R$ and $T$ respectively and $\Theta_{\mu\nu}$ is defined as 
\begin{equation}
\Theta_{\mu\nu} \equiv g^{\alpha\beta}\frac{\delta T_{\alpha\beta}}{\delta g^{\mu\nu}}=-2T_{\mu\nu} +g_{\mu\nu}L_m -2g^{\alpha\beta}\frac{\partial^2 L_m}{\partial g^{\mu\nu}\partial g^{\alpha \beta}}.
\end{equation}
To simplify highly nonlinear field equation we require  a specific form of $f(R,T)$. In this paper we consider $f(R, T)= R+2f(T)$, where $f(T)$ is an arbitrary function of the trace of the energy-momentum tensor of matter. The term $f(T)$ modifies the gravitational interaction between matter and curvature. For the  particular choice $f(R, T)= R+2f(T)$, gravitational field equation yields 
\begin{equation}
 R_{\mu\nu}-\frac{1}{2}R g_{\mu\nu} =T_{\mu\nu}+f(T) g_{\mu\nu}-2 f^\dagger(T) \left[T_{\mu\nu} + \Theta_{\mu\nu} \right],
\end{equation} 
where $f^\dagger(T)$ denotes the derivative of $f(T)$ with respect to $T$. 
We consider the  flat homogeneous and isotropic space-time given by Friedmann-Robertson-Walker (FRW) metric
\begin{equation}
ds^2=-dt^2+a^2 (t) \left[dr^2+ r^2 (d\theta^2+sin^2\theta d\phi^2 ) \right] ,                       
\end{equation}     
where $a(t)$ is the scale factor of the universe. In the present study we define the energy-momentum tensor of the metric as given by 
\begin{equation}
T_{\mu\nu}=(\rho+ \bar{p})u_{\mu} u_{\nu} - \bar{p} g_{\mu\nu} ,
\end{equation}
where $\rho$ is energy density of the universe, $\bar{p}$ is the effective pressure, $u^{\mu}$ is the four velocity and  $u^{\mu}u_{\mu}=1$. The matter Lagrangian density can be taken as $L_m =-\bar{p}$ and the trace of the total energy momentum tensor is given by $T=\rho-3\bar{p}$.  The expression of  $\Theta_{\mu\nu}$ is given by 
\begin{equation}
 \Theta_{\mu\nu}= -2 T_{\mu\nu}- \bar{p}\;g_{\mu\nu}.
\end{equation}
 Using eqs. (7) and (4), the gravitational field equations are given by 
  \begin{equation}
 R_{\mu\nu}-\frac{1}{2}R g_{\mu\nu} =T_{\mu\nu}+2f^\dagger(T) T_{\mu\nu}+\left[2 \bar{p} f^\dagger(T)+ f(T)\right] g_{\mu\nu} .
\end{equation}
 The reconstruction of arbitrary FRW cosmologies is possible by an appropriate choice \cite{harko} of the function $f(T)$. The simplest cosmological model can be obtained by choosing the function  $f(T)$     so that   $f(T) = \lambda T $, where $\lambda $  is  a  constant acting as a coupling parameter between geometry and matter. The field equations for a particular choice  of the function $f(T)=\lambda T$, where $\lambda $  is an arbitrary constant,  yield 
\begin{equation}
3H^2 = (1+ 3\lambda) \rho -\lambda \bar{p},
\end{equation} 
\begin{equation}
 2\dot{H} + 3H^2 =\lambda \rho -( 1+ 3\lambda) \bar{p} ,      
\end{equation}                   
where $ H=\frac{\dot{a}}{a} $ is the Hubble parameter and an  over-dot represents derivative with respect to cosmic time ($t$). Let the effective pressure  ($\bar{p}$) contain two parts : $\bar{p}=p+\Pi$, where $p$ is the isotropic pressure of the universe and   $\Pi \;(\leq 0) $ is the bulk viscous pressure. Here we consider linear equation of state (EoS) of the cosmological  fluid, i.e., 
\begin{equation}
p=\omega \rho ,  
\end{equation}
where $\omega \;(1\geq\omega\geq 0) $ represents EoS parameter. Using Eqs. (9)-(11), we obtain respective expressions of energy density and bulk viscous pressure, which are given by  
 \begin{equation}
\rho= \frac{3}{1+4\lambda} H^2- \frac{2\lambda }{(1+2\lambda)(1 + 4\lambda)}\dot{H} , 
\end{equation}
 \begin{equation}
\Pi=-\frac{3(1+\omega)}{1+4\lambda} H^2 -\frac{2(1+3\lambda-\omega\lambda)}{(1+2\lambda)(1 + 4\lambda)}  \dot{H}  . 
\end{equation}
It is worthy to notice that one can recover the standard cosmological models with viscosity \cite{rev13,rev14,zimdahl} from field Eqs. (12)-(13) for $\lambda =0$. For physically viable cosmological solutions in the presence of bulk viscosity one requires $\Pi<0$ and $|\Pi|<< \rho$.  In the following section, we shall study cosmological solution in the presence of bulk viscosity and matter described by the linear EoS in the $f(R,T)$ theory of gravity. Different phases of evolution of the universe can be studied by any  relevant cosmological quantity like deceleration parameter. The deceleration parameter ($q$) is related to Hubble parameter ($H$) as $$q=\frac{d}{dt}\left(\frac{1}{H}\right)-1 .$$ The accelerating phases of evolution of the universe are obtained for negative values of the deceleration parameter ($q<0$),  whereas the  decelerating phases of evolution of the universe are obtained for positive values of the deceleration parameter ($q>0$). 
\section{Cosmological  Solutions}
 The bulk viscous stress satisfies following transport equation \cite{eckart, israel}
\begin{equation}
\Pi +\tau\dot{\Pi}=-3\zeta H-\frac{\epsilon}{2}\tau\Pi\left(3H+\frac{\dot{\tau}}{\tau}-\frac{\dot{\zeta}}{\zeta}-\frac{\dot{T}}{T} \right) ,            
\end{equation}
where the parameter $\zeta (\geq 0)$ is the co-efficient of bulk viscosity, the parameter $\tau (\geq 0)$ is the relaxation time and the constant $\epsilon$ has two values, either 0 or 1. The transport Eq. (14) reduces to Full Israel Stewart (FIS) theory for $\epsilon =1$, for $\epsilon =0$ it reduces to  Truncated Israel Stewart (TIS). One can recover Eckart theory for  $\tau=0$.  The positive entropy production due to bulk viscosity is confirmed by the positive values of the co-efficient of bulk viscosity ($\zeta$).
The set of Eqs. (9)-(14) are employed to obtain cosmological solutions. The systems of equations are not closed as the number of equations is  less than the number of unknowns.  It is known that the coefficient of bulk viscosity  and relaxation time are, in general, functions of time (or of the energy density). We, therefore, consider following relation \cite{brevik,meng,jou}
\begin{equation}
\zeta=\beta\;\rho^s ,\;\;\;\; \tau=\beta \;\rho^{s-1} ,
\end{equation}            
where   $\beta\;(>0)$ and $s\;(>0)$ are  constants. It could be mentioned here that recent analysis of B. D. Normann and I. Brevik \cite{rev20,rev21} tend to favor the choice $s=\frac{1}{2}$. 
\subsection{Eckart Theory ($\tau=0$):} Using Eqs. (9)-(15), for $\tau=0$ we get
\begin{equation}
\frac{2(1+3\lambda-\omega\lambda)}{(1+2\lambda)(1 + 4\lambda)}  \dot{H} +\frac{3(1+\omega)}{1+4\lambda} H^2 
 = 3\beta \left[\frac{3H^2}{1+4\lambda} - \frac{2\lambda \dot{H} }{(1+2\lambda)(1 + 4\lambda)} \right]^s H. 
\end{equation}                
The evolution of the universe in Eckart theory can be obtained by using Eq. (16) which is highly nonlinear to obtain a general analytic solution. However, using  Eq. (16), one can obtain numerical solutions in term of cosmic time of  relevant parameter, such as  deceleration parameter ($q$), Hubble parameter ($H$) and  scale factor ($a$)   to study  different phases of the universe studying in Eckart theory. The late-time behavior of the models is better revealed from the plot of deceleration parameter ($q$) vs redshift parameter ($z$).  The Hubble parameter of the universe is related to the redshift parameter as ($H=-\frac{1}{1+z}\frac{dz}{dt}$). To study late-time evolution of the universe, we rewrite Eq. (16)  in terms of deceleration parameter $(q)$ and redshift parameter ($z$), which yields
$$
\frac{2(q+1)(1+3\lambda-\omega\lambda)}{1+2\lambda}-3(1+\omega)+\frac{3\beta (\frac{2}{3}H_1)^{2s-1}}{(1+4\lambda)^{s-1}}\left[3+\frac{2\lambda(q+1)}{(1+2\lambda)}\right]^s (1+z)^{\frac{3(2s-1)}{2}}=0.
$$ Here cosmic time ($t$) is related to redshift parameter ($z$) by the relation $t=\frac{H_1^{-1}}{(1+z)^{\frac{3}{2}}}$ \cite{Sch}, where the  constant $H_1$ has  unit (Gyr)$^{-1}$. The lower values of the redshift parameter ($z$) indicates late-time behavior of the universe and higher values of the redshift parameter ($z$) indicates early-time behavior of the universe. Figure (1)  shows the plot of $q$ vs $z$ for different values of $\lambda$ for a given set of other parameters.  It indicates that the values of the deceleration  parameter ($q$) are lower for higher values of the coupling parameter $\lambda$ at given instant of redshift parameter ($z$). For higher values of coupling parameter ($\lambda$) the universe enters to the late-time acceleration phase earlier.  The plot of scale factor $a(z)$ vs redshift parameter $(z)$ in Eckart theory for different values of $\lambda$ for a given set of other parameters, as in Fig (2),  indicates that the scale factor of the universe increases more rapidly  for higher values of coupling parameter $(\lambda)$. As the above Eq. (16) is highly non-linear and the relativistic solution cannot be expressed in known general analytic form, we consider special cases for simplicity.
 \begin{figure}[ph]
\centerline{\psfig{file=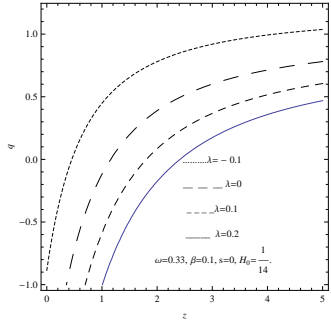,width=1.7in}}
\vspace*{8pt}
\caption{shows the plot of   $q$  vs  $z$ for Eckart theory in $f(R,T)$ theory of  gravity for different values of $\lambda$ for a given set of other parameters. \label{fig1}}
\end{figure}
\begin{figure}[ph]
\centerline{\psfig{file=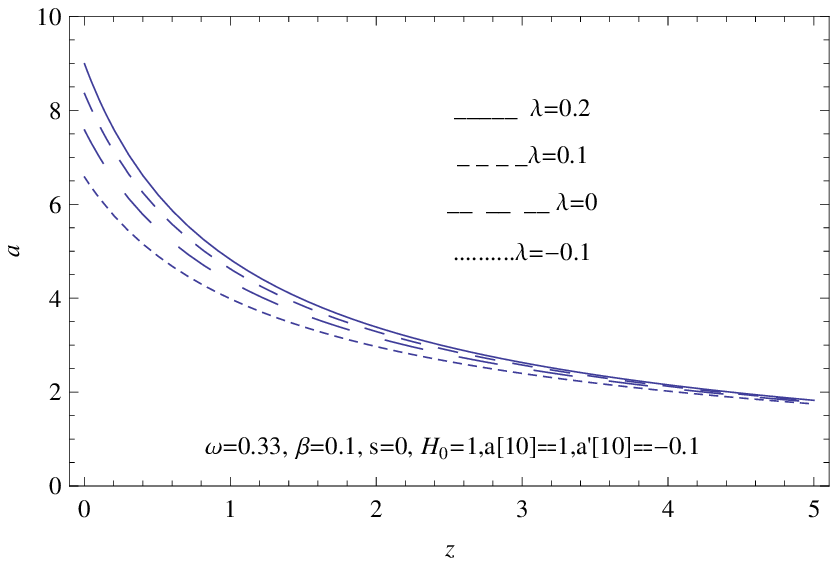,width=1.7in}}
\vspace*{8pt}
\caption{shows the plot of   $a$  vs  $z$ for Eckart theory in $f(R,T)$ theory of gravity for different values of $\lambda$ for a given set of other parameters. \label{fig2}}
\end{figure}
\subsubsection{Power law model :} In this particular case scale factor of the universe exhibits power law expansion
$a(t)=a_0 t^D$, where $a_0$ and $D$ are constants. For power law model, the expression of energy density and bulk viscosity stress are given by 
\begin{equation}
\rho=\rho_0 t^{-2},\;\;\;\;  \Pi=-\Pi_0 t^{-2} ,
\end{equation}
where $\rho_0=\frac{3D+6\lambda D+2\lambda }{(1+4\lambda)(1+2\lambda)}D$ and $\Pi_0 = \left[\frac{3D(1+\omega)(1+2\lambda)-2(1+3\lambda-\omega\lambda)}{(1+4\lambda)(1+2\lambda)} \right] D  $. For physically permissible  (i.e., $\Pi < 0$) solutions we obtain following lower boundary limit of power law exponent  $D>\frac{2(1+3\lambda -\omega \lambda)}{(1+\omega)(1+2\lambda)}$.
 For power law expansion Eq. (16) becomes 
\begin{equation}
A_1 + A_2 t^{1-2s}=0,
\end{equation}
 where $A_2=3\beta(1+4\lambda)^{1-s}\left[3D^2+\frac{2\lambda D}{1+2\lambda}\right]^s $ and $A_1=2-\frac{3(1+2\lambda)(1+\omega)D}{1+3\lambda-\omega \lambda}$. For $s\neq \frac{1}{2}$ Eq. (18) yields $A_1=A_2=0,$ i.e., $D=\frac{2(1+3\lambda-\omega\lambda)}{3(1+\omega)(1+2\lambda)}$ and $\lambda=-\frac{1}{4}$ or $\beta=0$ which is not physically acceptable. For $s=\frac{1}{2}$, Eq. (18) yields $A_1 + A_2$=0, which leads to 
\begin{equation}
3\beta\frac{(1+2\lambda)\sqrt{(1+4\lambda)}}{1+3\lambda-\omega\lambda}\left[3D^2+\frac{2\lambda D}{1+2\lambda}\right]^{\frac{1}{2}} + 
2-\frac{3(1+2\lambda)(1+\omega)D}{1+3\lambda-\omega \lambda}=0.
\end{equation}
\begin{figure}[ph]
\centerline{\psfig{file=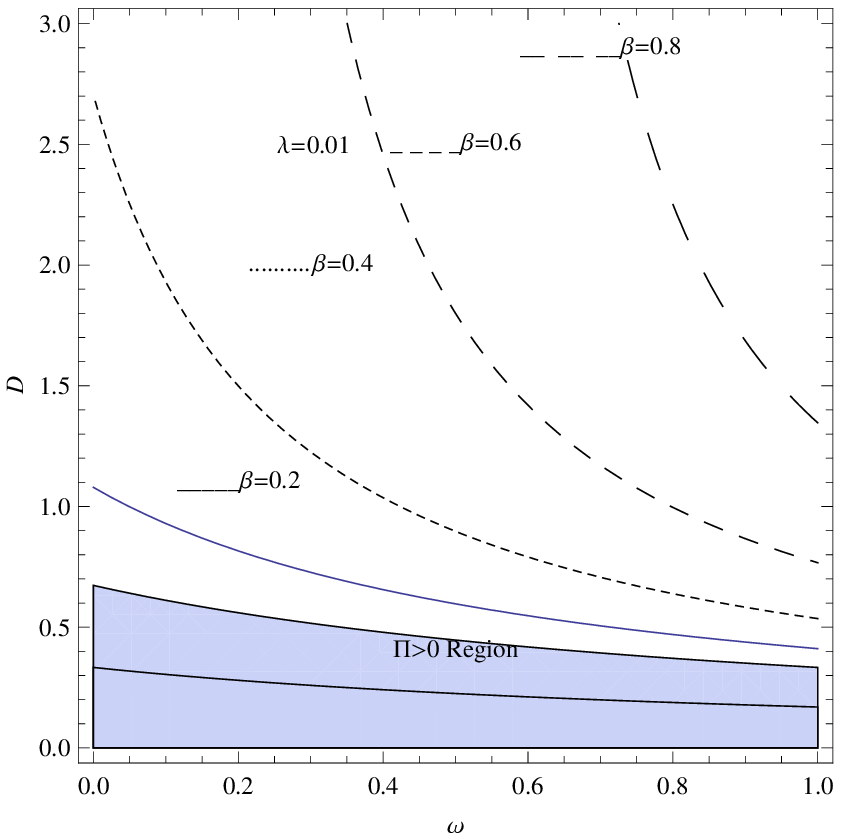,width=1.7in}}
\vspace*{8pt}
\caption{shows the plot of   $D$  vs  $\omega$ for power-law evaluation for different values of $\beta$ in Eckart theory of $f(R,T)$ gravity with $\lambda=0.01$. \label{fig3}}
\end{figure}
Using Eq. (19) a power law accelerated universe $D>1$ is obtained in Eckart theory for different values of other parameters as shown in Fig. (3). In the Fig. (3), the shadow region is physically unsuitable for power law expansion  as it permits $\Pi>0$ and $|\Pi| >\rho$. Figure (3) shows that for power law evolution the exponent ($D$) is higher for smaller value of EoS parameter ($\omega$) and higher value of bulk viscous constant ($\beta$) for a given value of other parameters in $f(R,T)$ theory of gravity. So power law acceleration is suitable for smaller values of EoS parameter ($\omega$) and higher values of bulk viscous constant ($\beta$). 
\subsubsection{Exponential model :}
 The exponential  expansion $(a(t)\sim \exp[H_0 t])$ of the universe with viscosity  may be obtained  in $f(R,T)$ gravity by setting $H=H_0=const.$ In this case the equation (16) yields 
$H_0=\left[\frac{\beta(1+4\lambda)^{1-s}}{(1+\omega)3^{-s}}\right]^{\frac{1}{1-2s}} .$ 
 So an exponential acceleration of the universe in $f(R,T)$ gravity is permitted  for coupling parameter $\lambda>-\frac{1}{4}$. The rate of exponential acceleration is higher for smaller values of EoS parameter ($\omega$) and higher values of  bulk viscous constant ($\beta$)  for $s<\frac{1}{2}$.  However for $s>\frac{1}{2}$, the rate of exponential acceleration is higher for higher values of EoS parameter ($\omega$) and smaller values of bulk viscous constant ($\beta$). In exponential evolution  the energy density ($\rho = const.$) and bulk viscous stress  remain constant ($\Pi= const.)$ parameters.
\subsection{ Truncated Israel Stewart Theory $(\epsilon =0)$ :}
 Using equations (9)-(15) for Truncated Israel Stewart (TIS) theory  ($\epsilon = 0$) in a flat universe, we obtain
\[ 
\ddot{H} +3\left[\frac{1+\omega+3\lambda+2\omega\lambda}{1+3\lambda-\omega\lambda} \right]H\dot{H} -\frac{9(1+2\lambda)}{2(1+3\lambda-\omega\lambda)} H^3+
\]
\begin{equation} 
 \frac{1}{\beta}\left[\frac{3(1+2\lambda)(1+\omega) H^2 }{2(1+3\lambda-\omega\lambda)}+\dot{H}\right] \left[\frac{3(1+2\lambda)H^2-2\lambda \dot{H}}{(1+2\lambda)(1 + 4\lambda)} \right]^{1-s} =0 .
\end{equation}                
  The evolution of the universe in TIS theory for linear EoS in $f(R,T)$ theory of gravity can be obtained by using above Eq. (20) which is highly nonlinear to obtain a general analytic  solution of known form of the scale factor of the universe. However, one can obtain numerical solution of the relevant parameter, such as deceleration parameter ($q$), the Hubble parameter ($H$) and  scale factor ($a(t)$) in term of cosmic time to study different phases of the universe in TIS theory by using Eq. (20). The late-time behavior of the models is better revealed from the plot of deceleration parameter ($q$) vs redshift parameter ($z$). To study late-time evolution of the universe we rewrite the above Eq. (20) in terms of the deceleration parameter $(q)$ and the redshift parameter ($z$), which yields 
	\[
	q^{\prime} +\frac{2(q+1)^2}{1+z}-\frac{3(1+\omega+3\lambda+2\omega \lambda)(1+q)}{(1+3\lambda-\omega \lambda)(1+z)} -\frac{9(1+2\lambda)}{2(1+3\lambda-\omega \lambda)(1+z)} +  \frac{(2H_1)^{1-2s}}{\beta 3^{1-2s}} \times
	\]
$$ \left[\frac{3(1+2\lambda)(1+\omega)}{2(1+3\lambda-\omega\lambda}-(q+1)\right] \left[ \frac{3(1+2\lambda)+2\lambda(q+1)}{(1+2\lambda)(1+4\lambda)}\right]^{1-s} (1+z)^{\frac{1}{2}-3s} =0 .
	$$
	Here prime ($'$) represents derivative with respect to the redshift parameter ($z$). Figure (4) shows the plot of deceleration parameter $(q)$ vs redshift parameter $(z)$ for different values of coupling parameter $\lambda$ for a given set of other parameters in TIS theory. 	
	Figure (4) indicates that the values of the deceleration  parameter ($q$) are lower for higher values of the coupling parameter $\lambda$ at given instant of redshift parameter ($z$). For higher values of coupling parameter ($\lambda$) the universe enters into the late-time acceleration phase earlier.  The plot of scale factor $a(z)$ vs redshift parameter $(z)$ in TIS theory for different values of $\lambda$ for a given set of other parameters (as in Fig (5)) indicates that scale factor of the universe increases more rapidly  for higher values of coupling parameter $(\lambda)$. 	 As the above Eq. (20) is highly non-linear and the relativistic solution cannot be expressed in known general analytic form, we consider following special cases for simplicity.  
\begin{figure}[ph]
\centerline{\psfig{file=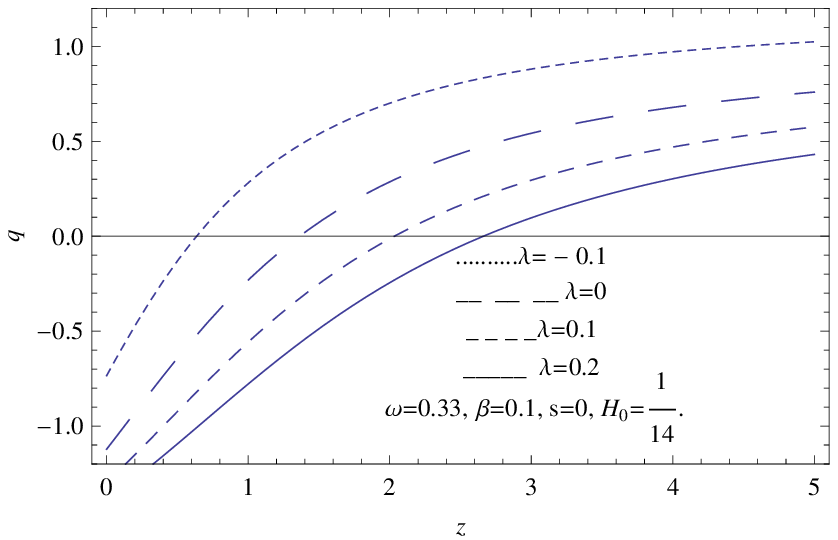,width=1.7in}}
\vspace*{8pt}
\caption{shows the plot of   $q$  vs  $z$ for TIS theory in $f(R,T)$ theory of gravity for different values of $\lambda$ for a given set of other parameters.\label{fig4}}
\end{figure}
\begin{figure}[ph]
\centerline{\psfig{file=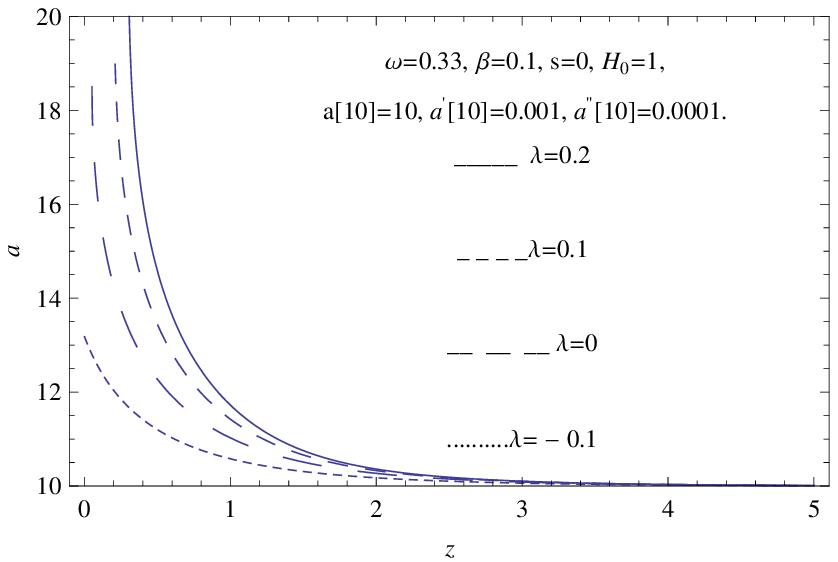,width=1.7in}}
\vspace*{8pt}
\caption{shows the plot of   $a$  vs  $z$ for TIS theory in $f(R,T)$ theory of gravity for different values of $\lambda$ for a given set of other parameters.\label{fig5}}
\end{figure}
\subsubsection{Power-law model :}
In the power-law expansion ($a(t) = a_0 t^D $, where $a_0$ and $D$ are constants) the expressions of energy density and bulk viscosity stress becomes 
\begin{equation}
\rho=\rho_0 t^{-2},\;\;\;\;  \Pi=-\Pi_0 t^{-2} ,
\end{equation}
where the constant  $\rho_0=\frac{3D+6\lambda D+2\lambda }{(1+4\lambda)(1+2\lambda)}D$ and the constant  $\Pi_0 = \left[\frac{3D(1+\omega)(1+2\lambda)-2(1+3\lambda-\omega\lambda)}{(1+4\lambda)(1+2\lambda)} \right] D  $. A physically viable  solution is permitted for  $\Pi < 0$ which follows lower boundary limit of power-law exponent  $D>\frac{2(1+3\lambda -\omega \lambda)}{(1+\omega)(1+2\lambda)}$.
For power law expansion  Eq. (20)  yields,
\begin{equation}
    B_1 + B_2 t^{2s-1} =0 ,
\end{equation}
where the constant $B_1 = (\frac{4(1+3\lambda-\omega\lambda)}{(1+2\lambda)}-6(1+\omega)D-\frac{6\lambda D}{1+2\lambda}-9D^2)$ 
and the constant $B_2 =(\frac{3(1+\omega)D}{\beta}-\frac{2(1+3\lambda-\omega\lambda)}{\beta(1+2\lambda)})(\frac{3D^2}{1+4\lambda}+\frac{2\lambda D}{(1+2\lambda)(1+4\lambda)})^{1-s}$. For $s\neq\frac{1}{2}$ Eq. (22) yields $B_1=B_2=0$, i.e., $D=\frac{2(1+3\lambda-\omega\lambda)}{3(1+2\lambda)(1+3\gamma)}$ and $\lambda=-\frac{1}{4}$ which is not physically acceptable. For  $s=\frac{1}{2}$, one obtains  $B_1 + B_2=0,$  which leads to
\[
4(1+3\lambda-\omega\lambda)-6(1+\omega)(1+2\lambda)D-6\lambda D-9(1+2\lambda)D^2 +\frac{1}{\sqrt{(1+4\lambda)}\beta}\times\] 
\begin{equation}
\left[3(1+\omega)(1+2\lambda)D-2(1+3\lambda-\omega\lambda)\right]\times
\left[3D^2+\frac{2\lambda D}{(1+2\lambda)}\right]^{\frac{1}{2}} =0 .
\end{equation}
 Using Eq. (23) a power law accelerated ($D>1$) universe is obtained in TIS theory for different values of other parameters as shown in Fig. (6). In the Fig. (6), the shadow region corresponds physically unacceptable solution  for power law expansion  as in this region $\Pi>0$ and $|\Pi| >\rho$. Figure (6) shows that for power law evolution the exponent ($D$) is higher for smaller values of EoS parameter ($\omega$) and higher values of bulk viscous constant ($\beta$) in $f(R,T)$ theory of  gravity for a given value of other parameters. So power law acceleration is suitable for smaller values of EoS parameter ($\omega$) and higher value of bulk viscous constant ($\beta$).  
\begin{figure}[ph]
\centerline{\psfig{file=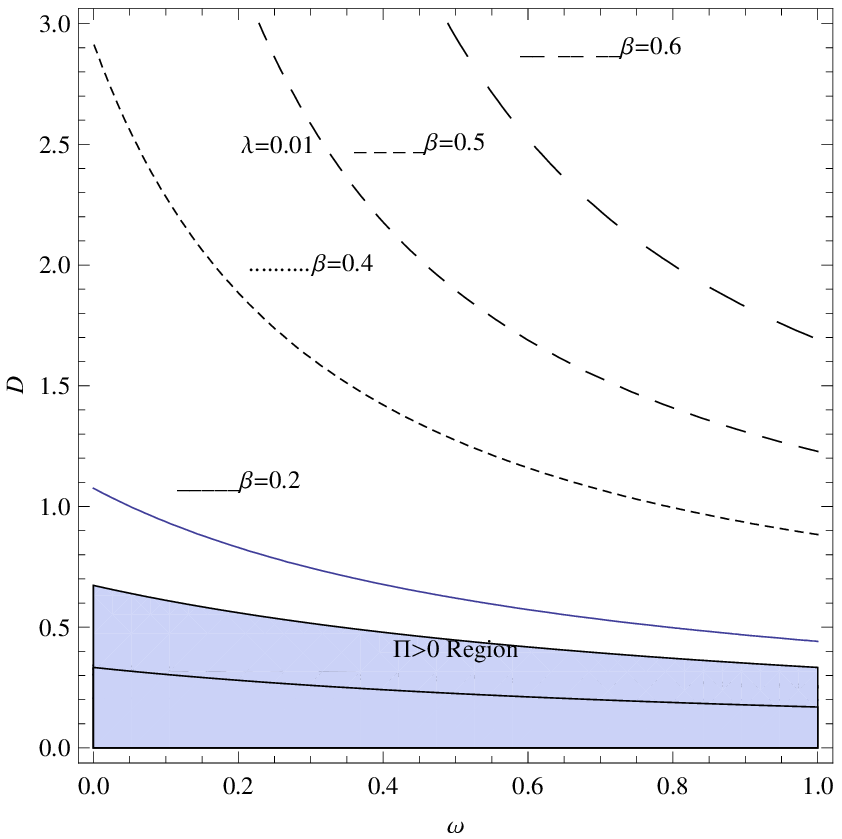,width=1.7in}}
\vspace*{8pt}
\caption{shows the plot of   $D$  vs  $\omega$ for Power-law evaluation in TIS theory for different values of $\beta$ in $f(R,T)$ gravity with $\lambda=0.01$.\label{fig6}}
\end{figure}
	\subsubsection{Exponential Model :}
 An exponential  evolution ($a(t)\sim \exp[H_0 t]$) of the universe in $f(R,T)$ gravity with TIS theory   may be obtained   by setting $H=H_0=const.$ For the exponential evolution  Eq. (20) yields 
\begin{equation}
H_0=\left[\frac{\beta(1+4\lambda)^{1-s}}{(1+\omega)3^{-s}}\right]^{\frac{1}{1-2s}} .
\end{equation} 
 The exponential accelerations of the universe in $f(R,T)$ gravity with viscosity, described by TIS theory, are permitted for the coupling parameter $\lambda>-\frac{1}{4}$.  The rate of exponential acceleration is higher for smaller values of EoS parameter ($\omega$) and higher values of bulk viscous constant ($\beta$)  for $s<\frac{1}{2}$.  However for $s>\frac{1}{2}$, the rate of exponential acceleration is higher for higher values of EoS parameter ($\omega$) and smaller values of bulk viscous constant ($\beta$). For exponential evolution in $f(R,T)$ theory of gravity, the energy density ($\rho = const.$) and bulk viscous stress ($\Pi= const.)$  remain constant parameters. 
\subsection{ Full Israel Stewart Theory $(\epsilon =1)$ :}
Using Eqs. (9)-(15) for Full Israel Stewart (FIS) theory  ($\epsilon = 1$) in a flat universe, we obtain the field equation
\[ 
\ddot{H} +\frac{1+\omega+3\lambda+2\omega\lambda}{1+3\lambda-\omega\lambda} 3 H \dot{H}-\frac{9(1+2\lambda)H^3}{2(1+3\lambda-\omega\lambda)}+\]
\[
\frac{1}{\beta}\left[ \frac{3(1+\omega)(1+2\lambda)H^2}{2(1+3\lambda-\omega\lambda)} +\dot{H}\right] \times\left[\frac{3H^2}{1+4\lambda}-\frac{2\lambda\dot{H}}{(1+2\lambda)(1+4\lambda)}\right]^{1-s}\]
\begin{equation}
+\frac{1}{2}\left[\frac{3(1+\omega)(1+\lambda)H^2}{2(1+3\lambda-\omega \lambda)} + \dot{H}\right] \times
\left[3H-\frac{1+2\omega}{1-2\omega}\frac{6(1+2\lambda)H\dot{H}-2\lambda \ddot{H}}{3(1+2\lambda)H^2-2\lambda\dot{H}} \right] =0 ,
\end{equation}                
  where we consider barotropic behavior of temperature ($T\sim \rho^{\frac{\omega}{1+\omega}}$). The evolution of the universe in FIS theory for linear EoS in $f(R,T)$ gravity can be obtained by using Eq. (25), which is highly nonlinear, to obtain a general analytic solution. However,  numerical solutions of relevant parameters such as the deceleration parameter $(q)$, the Hubble parameter $(H)$ and  the scale factor $(a)$ can be obtained  for a given set of other parameters to study evolutions of the universe. The late-time behavior of the models is better revealed from the plot of deceleration parameter ($q$) vs redshift parameter ($z$). To study late-time evolution of the universe we rewrite the above Eq. (25) in terms of the deceleration parameter $(q)$ and the redshift parameter ($z$), which yields   
		\[
	q^{\prime} +\frac{2(q+1)^2}{1+z}-\frac{3(1+\omega+3\lambda+2\omega \lambda)(1+q)}{(1+3\lambda-\omega \lambda)(1+z)} -\frac{9(1+2\lambda)}{2(1+3\lambda-\omega \lambda)(1+z)} +  \frac{(2H_1)^{1-2s}}{\beta 3^{1-2s}} \times
	\]
\[ \left[\frac{3(1+2\lambda)(1+\omega)}{2(1+3\lambda-\omega\lambda}-(q+1)\right] \left[ \frac{3(1+2\lambda)+2\lambda(q+1)}{(1+2\lambda)(1+4\lambda)}\right]^{1-s} (1+z)^{\frac{1}{2}-3s}+\frac{1}{2(1+z)} \times  
\]
$$\left[\frac{3(1+\omega)(1+\lambda)}{2(1+3\lambda-\omega\lambda)}-1-q\right]\left[3+\frac{1+2\omega}{1-2\omega}(2(q+1)+\frac{2\lambda q^{\prime} (1+z)}{3(1+2\lambda)+2\lambda (q+1)})\right]=0.
	$$
Figure (7) shows the plot of deceleration parameter $(q)$ vs redshift parameter $(z)$ for different values of coupling parameter $\lambda$ for a given set of other parameters in FIS theory. 	
	Figure (7) indicates that the values of the deceleration  parameter ($q$) are lower for higher values of the coupling parameter $\lambda$ at given instant of redshift parameter ($z$). For higher values of coupling parameter ($\lambda$) the universe enters into the late-time acceleration phase earlier.  The plot of scale factor $a(z)$ vs redshift parameter $(z)$ in TIS theory for different values of $\lambda$ for a given set of other parameters as in Fig (8)  indicates that scale factor of the universe increases more rapidly  for higher value of coupling parameter $(\lambda)$.  As the above Eq. (25) is highly non-linear and the relativistic general analytic solution cannot be expressed in known form, we consider following special cases for simplicity. 
\begin{figure}[ph]
\centerline{\psfig{file=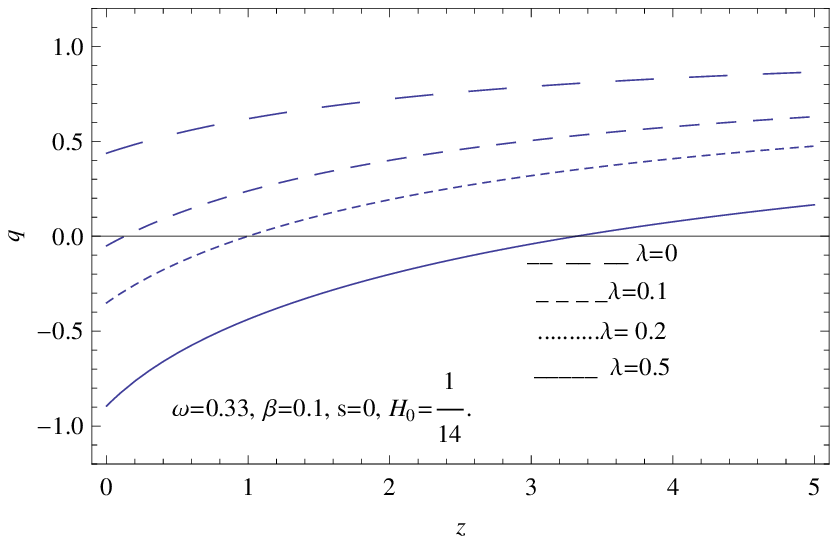,width=1.7in}}
\vspace*{8pt}
\caption{shows the plot of   $q$  vs  $z$ for FIS theory in $f(R,T)$ theory of gravity for different values of $\lambda$ for a given set of other parameters.\label{fig7}}
\end{figure}
\begin{figure}[ph]
\centerline{\psfig{file=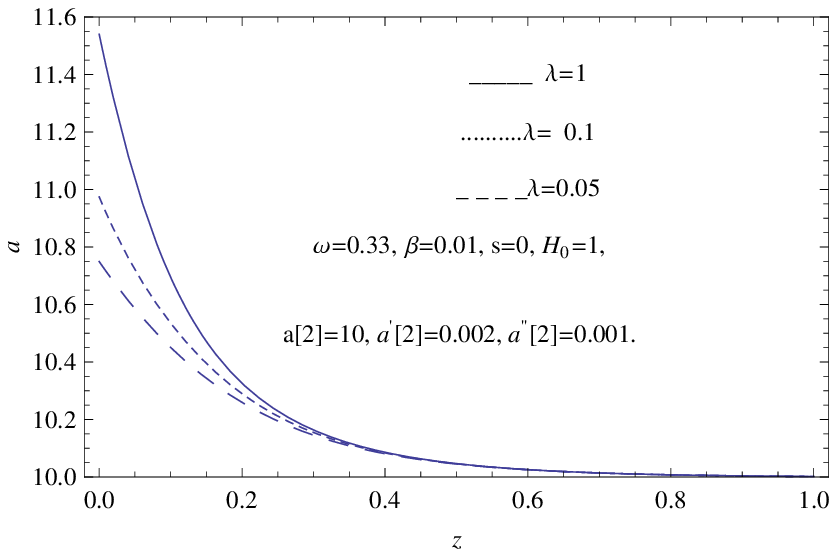,width=1.7in}}
\vspace*{8pt}
\caption{shows the plot of   $a$  vs  $z$ for FIS theory in $f(R,T)$ theory of gravity for different values of $\lambda$ for a given set of other parameters.\label{fig8}}
\end{figure}
\subsubsection{Power law model :}
We can obtain the power-law expansion of the scale factor $(a(t))$ of the universe in $f(R,T)$ gravity with FIS theory by setting $a(t) = a_0 t^D $, where $a_0$ and $D$ are constants. In power law model the expressions for energy density and bulk viscosity stress become 
\begin{equation}
\rho=\rho_0 t^{-2},\;\;\;\;  \Pi=-\Pi_0 t^{-2}
\end{equation}
where the constant $\rho_0=\frac{3D+6\lambda D+2\lambda }{(1+4\lambda)(1+2\lambda)}D$ and the constant $\Pi_0 = \left[\frac{3D(1+\omega)(1+2\lambda)-2(1+3\lambda-\omega\lambda)}{(1+4\lambda)(1+2\lambda)} \right] D  $. For  physically permitted solutions  $\Pi < 0$ which shows the following lower boundary limit of  power law exponent  $D>\frac{2(1+3\lambda -\omega \lambda)}{(1+\omega)(1+2\lambda)}$. 
For power law expansion  Eq. (25)  yields,
\begin{equation}
    C_1 + C_2 t^{2s-1} =0 ,
\end{equation}
where  the constant $C_1 =2-3D (\frac{1+\omega+3\lambda+2\omega\lambda}{1+3\lambda-\omega\lambda})-\frac{9D^2 (1+2\lambda)}{2(1-3\lambda -\omega\lambda)}+\frac{D}{2}(\frac{3(1+\omega)(1+2\lambda)D}{2(1+3\lambda-\omega\lambda)}-1)(3+\frac{(1+2\omega)(6(1+2\lambda D)+2\lambda)}{(1+\omega)(3(1+2\lambda D)+2\lambda)}) $ 
and the constant  $C_2 =(\frac{3(1+\omega)(1+2\lambda)D}{2\beta(1+3\lambda-\omega\lambda)}-\frac{1}{\beta})(\frac{3D^2}{1+4\lambda}+\frac{2\lambda D}{(1+2\lambda)(1+4\lambda)})^{1-s}$. For $s\neq\frac{1}{2}$ Eq. (27) yields $C_1=C_2=0$, i.e., $D=\frac{2(1+3\lambda-\omega\lambda)}{3(1+2\lambda)(1+3\gamma)}$ and $\lambda=-\frac{1}{4}$ which is not physically acceptable. For  $s=\frac{1}{2}$, one obtains  $C_1 + C_2=0,$  which leads to
\[
2-3D \left[\frac{1+\omega+3\lambda+2\omega\lambda}{1+3\lambda-\omega\lambda}\right]-\frac{9D^2 (1+2\lambda)}{2(1-3\lambda -\omega\lambda)} +\frac{D}{2}\times
\]
\[
\left[\frac{3(1+\omega)(1+2\lambda)D}{2(1+3\lambda-\omega\lambda)}-1\right]\times\left[3+\frac{(1+2\omega)(6(1+2\lambda D)+2\lambda)}{(1+\omega)(3(1+2\lambda D)+2\lambda)}\right] \] 
\begin{equation}
+\left[\frac{3(1+\omega)(1+2\lambda)D}{2\beta(1+3\lambda-\omega\lambda)}-\frac{1}{\beta}\right] \times \left[\frac{D}{{1+4\lambda}}(3D+\frac{2\lambda}{(1+2\lambda)})\right]^{\frac{1}{2}}=0 .
\end{equation}
 Using Eq. (28) we can obtain power-law type accelerated  ($D>1$) evolution for a given set of other   parameters in FIS theory as shown in Fig. (9).   It is shown in Fig. (9) that the  shadow  regions  are  unsuitable  for power law evolution as it corresponds to $\Pi> 0$ and $|\Pi| >\rho$.
\begin{figure}[ph]
\centerline{\psfig{file=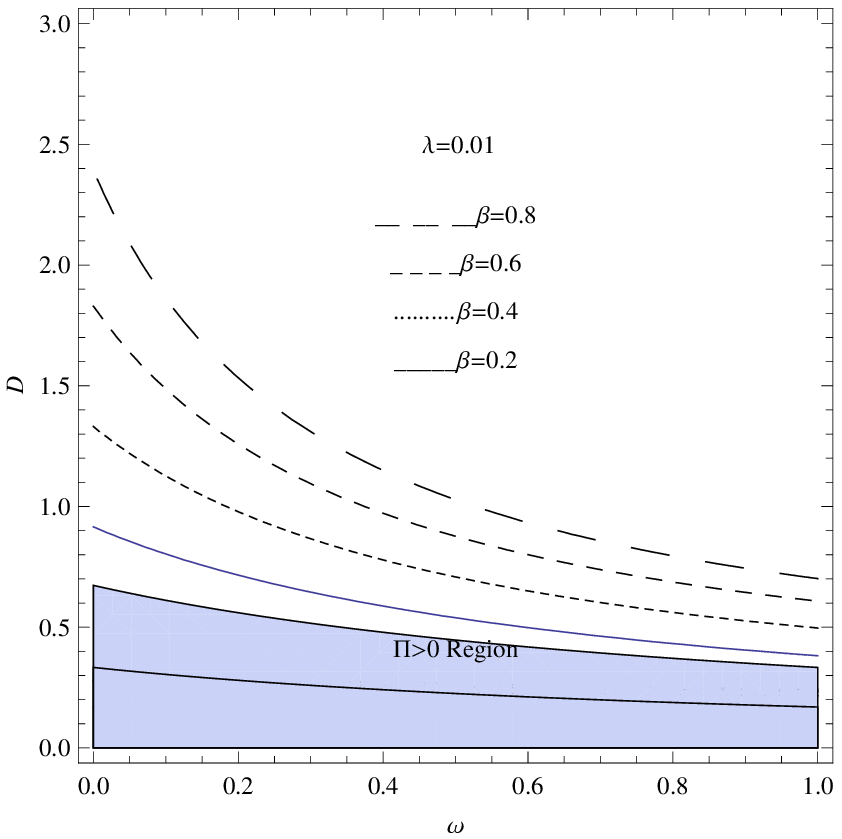,width=1.7in}}
\vspace*{8pt}
\caption{shows the plot of   $D$  vs  $\omega$ for Power-law evaluation in FIS theory for different values of $\beta$ in $f(R,T)$ gravity with $\lambda=0.01$.\label{fig9}}
\end{figure}
  Power law exponent $D$  is higher for smaller values of EoS parameter ($\omega$) and higher values of bulk viscous constant ($\beta$) for  given values of other parameter as shown in Fig. (9). Power law acceleration is suitable for smaller values of EoS parameter ($\omega$) and higher values of bulk viscous constant ($\beta$).  
	\subsubsection{Exponential Model :}
 The exponential  evolution  with viscosity described by FIS theory may be obtained  in $f(R,T)$ gravity by setting $H=H_0=const.$  In this special case the energy density ($\rho = const.$) and bulk viscous stress ($\Pi= const.)$ remain constants. For exponential expansion the Eq. (25) yields 
\begin{equation}
H_0=\left[\frac{\beta(1-\omega)(1+4\lambda)^{1-s}}{2(1+\omega)3^{-s}}\right]^{\frac{1}{1-2s}} .
\end{equation} 
 An exponential acceleration  of the universe is permitted  in $f(R,T)$ gravity of  FIS theory  for coupling parameter $\lambda>-\frac{1}{4}$ and EoS parameter $\omega\neq 1$.  The rate of exponential acceleration is higher for smaller values of EoS parameter ($\omega$) and higher values of bulk viscous constant ($\beta$)  for $s<\frac{1}{2}$.  However for $s>\frac{1}{2}$, the rate of exponential acceleration is higher for higher values of EoS parameter ($\omega$) and smaller values of bulk viscous constant ($\beta$).
\section{Constraints  on models' parameters from observational data :}
To find out constraints among the different parameters ($a_0,\; H_0,D,\;\lambda$), we will fit the  models with Redshift-Magnitude Observation from New, Old and Combined Supernova Data Sets \cite{kowalski}. The scale factor of the universe is related to redshift parameter $(z)$ as $\frac{a(t)}{a(t_0)}=\frac{1}{1+z}$, where $a(t_0)$ is a constant which may be considered  unity as a present value. For simplicity we consider power law and exponential models of the universe. 
\subsection{Power law model:}
 In power law model ($a(t)=a_0 t^D$)  the luminous modulus  is given by 
\begin{equation}
\mu(z) = 5\log\left(\left(\frac{1}{a_0}\right)^\frac{1}{D}\left(\frac{1+z}{D-1}\right)\left((1+z)^{\frac{D-1}{D}}-1\right)\right) +25 ,
\end{equation}
where $D\neq 1$. For $D=1$ i.e., neither acceleration or deceleration phase of evolution of the universe the luminous modulus  is given by 
$\mu=5\log\left(\frac{1+z}{a_0}\ln(1+z)\right) +25$. Using  Eq. (30) with observational data  we estimate the  constraints among the parameters of the models in $f(R,T)$ theory of gravity. 
Equation (30) is used to fit with recent observational data \cite{kowalski}.  The plot of $z$ vs $\mu(z)$ for power law evolution of the universe is shown  in Fig. (10). For power law evolution, it is evident that there is reasonable fit with observational data for $ 2.2>D> 1.9 $ with $a_0 = 1\times 10^{-8}$.
\begin{figure}[ph]
\centerline{\psfig{file=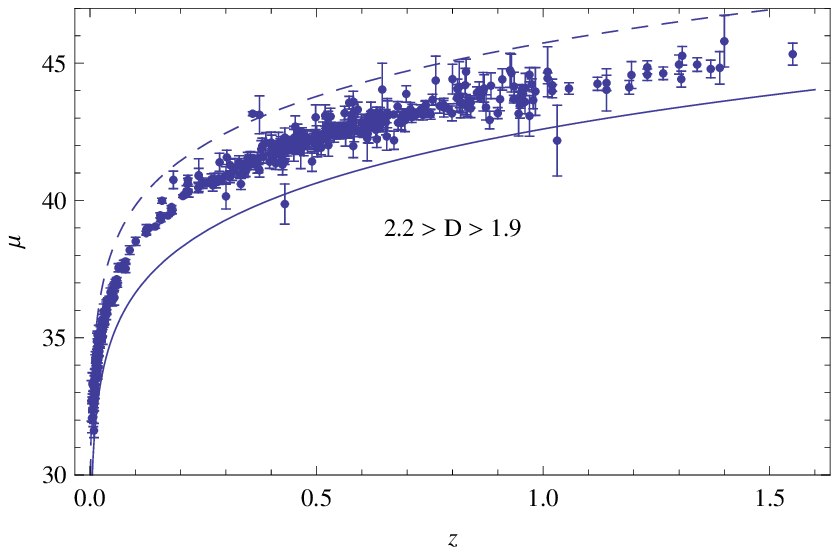,width=1.7in}}
\vspace*{8pt}
\caption{shows the plot of   $\mu$  vs  $z$ for supernova data  and power law evolution of the universe  for  $ D=2.2$ (solid line) and $D=1.9$ (dashed line) with $a_0 = 1\times 10^{-8}$.\label{fig10}}
\end{figure}
\subsection{Exponential model:}
 In exponential model, the scale factor of the Universe yields $a(t) \sim e^{H_0 t}$, where $H_0$ is a constant parameter. The expression of luminous modulus for the Universe in  this case is given by 
\begin{equation}
\mu(z) = 5\log\left(\frac{z(1+z)}{H_0}\right) +25 .
\end{equation}
Equation (31) will be useful now to fit with observational data \cite{kowalski}. The plot of $z$ vs $\mu(z)$, for exponential evolution to determine  constrains for relevant parameter, is shown in Fig. (11). For exponential model, it is evident that  the models fit well with observational data  for $0.5\times 10^{-3} >H_0>0.15\times 10^{-3}$, where $H_0$ has unit $(Gyr)^{-1}$.
\begin{figure}[ph]

\centerline{\psfig{file=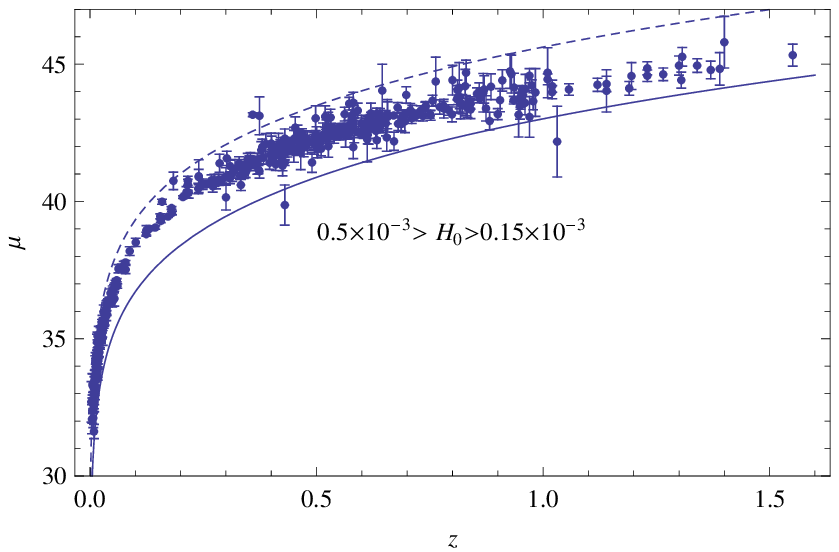,width=1.7in}}
\vspace*{8pt}
\caption{shows the plot of  $\mu$  vs  $z$ for supernova data and exponential evolution of the Universe  with $H_0 = 0.5\times 10^{-3}$(solid line) and $H_0 = 0.15\times 10^{-3}$ (dotted line).\label{fig11}}
\end{figure}
 \section{Conclusion}
In this paper an implementation of bulk viscous universe  model in the framework of $f(R,T)$ gravity with isotropic cosmological fluid described by linear equation of state ($p=\omega \rho$) has been done.  Bulk viscosity described by  Eckart, Truncated Israel Stewart (FIS) and Full  Israel Stewart (FIS) theories are considered  here to study geometrical and physical features of the universe in $f(R,T)=R+2\lambda T$ theory of gravity, where $\lambda$ is a constant acting as a coupling parameter between matter and geometry.  To find cosmological solutions,  special cases are considered for  non-linearity of field equations. We have studied numerical solutions, power law solutions and exponential solutions in the above mentioned viscous theories. Numerical solutions for the evolution of the universe are studied by considering cosmological parameters such as deceleration parameter $(q)$ and scale factor $(a)$. To study late-time behavior of the universe several plots of deceleration parameter ($q$) vs redshift parameter ($z$) are given as shown  in Fig (1), Fig. (4) and Fig. (7) for a given set of other parameters.  The figures show how deceleration parameter depends on redshift parameter for different values of coupling parameter ($\lambda$).  The figures also show late-time acceleration is permitted in Eckart, TIS and FIS theory. For larger values of $\lambda$ the universe enters into late-time acceleration  phase earlier in Eckart, TIS and FIS theory. To study late-time behavior of the universe several plots of scale factor ($a$) vs redshift parameter ($z$) are given  in Fig (2), Fig. (5) and Fig. (8) for a given set of other parameters. The scale factor increases more rapidly for  larger value of $\lambda$ at a given redshift parameter as shown in Fig (2), Fig. (5) and Fig. (8) in Eckart, TIS and FIS theory respectively. In the power law solution $(a(t)\sim t^D)$, the value of exponent $(D)$ is higher for higher values of bulk viscous constant $(\beta)$ and smaller values of EoS parameter $(\omega)$ as shown in Fig. (3), Fig (6) and Fig. (9) for Eckart, TIS and FIS theory respectively in $f(R,T)$ gravity.  Power law acceleration is suitable for smaller values of EoS parameter ($\omega$) and higher values of bulk viscous constant ($\beta$).   The exponential evolution of the universe in $f(R,T)$ gravity with Eckart, TIS and FIS theory is obtained for coupling parameter $\lambda>-\frac{1}{4}$ with constant energy density and bulk viscous stress.    
Redshift-Magnitude observations from supernova data are fitted  with the theoretical models to obtain best fit value of the different parameters as shown in Figs. (10)-(11). The power model fits with observational data for the values of power law exponent ($D$) in the boundary range $2.2 > D > 1.9$.  On the other hand, the exponential  model fits with observational data for the values of $H_0$,  is the range  $0.5\times 10^{-3} > H_0 > 0.15 \times 10^{-3}$, here the unit of $H_0$ is $(Gyr)^{-1}$.\\
In conclusion, it has been observed that the scale factor increases rapidly in all viscous theories i.e., Eckart, TIS and FIS theories in $f(R,T)$ gravity theory as compared to the respective all viscous theories in standard models for any positive values of coupling parameter $\lambda$. It is also noted that for larger values of coupling parameter $\lambda$ the universe enter into the late-time acceleration phase  earlier. 
\section*{Acknowledgement}
Author would like to thank the IUCAA Reference Centre at North Bengal University for extending necessary research facilities to initiate the work.  He would also like to thank unknown reviewers for their constructive critical comments to  improve upon the quality of the article.

\end{document}